\documentclass[pra,aps,twocolumn,superscriptaddress,showpacs,showkeys]{revtex4}
\usepackage{graphicx}
\usepackage{amsmath}
\usepackage{color}
\usepackage{amssymb}

\begin{document}

\title{ Optimum Phase Space Probabilities From Quantum Tomography}

\author{Arunabha S. Roy}
\affiliation{King's College, London}

\author{S. M. Roy}
\affiliation{HBCSE,Tata Institute of Fundamental Research, Mumbai}

\date{\today}

\begin{abstract}

We determine a positive normalised phase space probability 
distribution $P$ with minimum mean square fractional deviation from the Wigner distribution $W$ .The 
minimum deviation, an invariant under phase space rotations, is a quantitative measure of 
the quantumness of the state.The positive distribution closest to $W$ will be useful in quantum mechanics and 
in time frequency analysis .



\end{abstract}

\maketitle



 {\bf 1. Quasi-probability distributions in Quantum Mechanics and Time Frequency Analysis}. 
The Wigner quasi-probability distribution $W$ \cite{Wigner},first proposed to calculate quantum corrections to 
thermodynamic equilibrium, is now widely used in quantum mechanics,  
statistical mechanics, and technological areas such as time-frequency analysis of signals in
electrical engineering and seismology \cite{Ville-Cohen}. The $W$ distribution and other 
quasi-probability distributions such as the Husimi $Q$ function \cite{Husimi-AK} ,the Glauber-Sudarshan 
$P$ function and their $s$-parametrized generalizations \cite{Glauber} can be obtained in quantum optics
by measuring probability distributions of quadrature phases and making an inverse Radon transform, 
i.e. quantum tomography \cite{Vogel}. 
  
The Wigner function has the unique distinction of being the quantum analogue of the 
classical Liouville phase space distribution since its marginals reproduce quantum probability 
densities of position coordinates $q_i$, momentum coordinates $p_i$ and indeed of quadrature phases 
$q_i \cos \theta _i + p_i \sin \theta _i $ for all $ \theta _i$ with $i$ taking $N$ values for a $2N$-dimensional 
phase space.In time frequency analysis too $W$ has the correct marginals reproducing energy densities in time 
or frequency. Unlike the classical Liouville 
density, $W$ cannot be interpreted as a joint probability density, because there are quantum states for 
which $W$ is not positive definite. Similarly in time-frequency analysis, $W$ has  marginals 
reproducing the energy densities in time or frequency but cannot be interpreted as their joint density ; 
for that one uses the positive definite 'Spectrogram' even though it does not have 
the correct marginals. In quantum mechanics, the main reason for the importance of the Husimi function $Q$ 
( a smeared W function) is that it is 
positive definite ; secondly,as shown by Braunstein, Caves and Milburn, it is the optimum of the distributions 
obtained in the Von-Neumann-Arthurs-Kelly model for joint measurement of position and momentum \cite{Husimi-AK}.

In 2-dimensional phase space ,the Husimi function for a quantum state $\psi $ is a particular smearing of the Wigner 
function $W_{ \psi}(q',p')$ which is explicitly positive definite,
\begin{eqnarray}
 P_H(q,p)= \frac{1} {2\pi}|\big (\psi _{b,q,p },\psi \big )|^2 \nonumber \\
= \int dq' dp' W_{ \psi}(q',p') W_{\psi _{b,q,p } }(q',p' ),
\end{eqnarray}
where ,
\begin{equation}
 W_{\psi _{b,q,p } }(q',p' )=\frac{1} {\pi} \exp {(-\frac{(q-q')^2 } {2 b^2 } -2b^2 (p-p')^2) }
\end{equation}
is the Wigner function for the minimum uncertainty state centered at position $q$, momentum $p$
\begin{equation}
 \psi _{b,q,p } (q')= \frac{\exp {(-\frac{(q-q')^2 } {4 b^2 } +ipq') } } {(2\pi)^{1/4} \sqrt{b} }.
\end{equation}
The Husimi $Q$ function is obtained from $P_H(q,p) $ if we choose $b^2=1/2$.The variances differ from the 
true quantum values $(\Delta q)^2, (\Delta p)^2 $,
\begin{eqnarray}
 (\Delta q)_H^2 =(\Delta q)^2 +b^2,\> (\Delta p)_H^2 =(\Delta p)^2 +\frac{1}{4b^2}.
\end{eqnarray}
Hence, marginals 
of the Husimi function differ from the corresponding quantum probability densities,
even when the Wigner function (which has the correct marginals) is positive definite.This suggests 
that a positive distribution closer to the Wigner function may exist also in cases where the Wigner function is 
not positive definite. The acute need for the best such distribution can be illustrated in a practical context. 

{\bf Need for an optimum positive joint density function}. 
We give one example in time frequency analysis, where there is a practical need for 
such a positive distribution in order to define the bandwidth at a given time. We need to define the 
expectation values of frequency $\omega$ and its square $\omega ^2$ at time $t$ ; this is done easily if there is 
a positive density function $P$ (e.g. see Cohen \cite{Ville-Cohen} ),
\begin{equation}
 \langle \omega \rangle _{t}=\frac {\int d\omega \omega P(t, \omega )}{ \int d\omega P(t, \omega )},\>
 \langle \omega ^2 \rangle _{t}=\frac {\int d\omega \omega ^2 P(t, \omega )}{ \int d\omega P(t, \omega )}.
\end{equation}
However if we substitute the Wigner function $W(t,\omega )$ in place of $P(t, \omega )$ we obtain an expression 
for the square of the bandwidth at time $t$, in terms of the amplitude $A(t)$ of the signal,
\begin{equation}
 \langle \omega ^2 \rangle _{t} -(\langle \omega \rangle _{t})^2 = (1/2)\big ( (\dot{A}(t)/A(t) )^2-
\ddot{A}(t)/A(t)  \big ),
\end{equation}
which is not positive definite since the second term on the right-hand side can be negative.Thus the 
Wigner function does not yield a reasonable definition of the instantaneous band-width. The 
Husimi function will give a positive definite answer; but that answer 
may not be reliable since its marginals differ from those of $W$ even when $W$ is positive definite.
In quantum mechanics, exactly the same mathematics demonstrates the difficulty of 
defining the conditional dispersion in momentum for a given position using the Wigner function. 
The basic need for a probability interpretation in quantum mechanics, and an energy density interpretation 
in time-frequency analysis motivate the variational problem seeking the best possible positive distribution.
The positive joint probability we find has immediate utility for quantum mechanics (especially quantum optics) and 
in time-frequency analysis (with obvious transcriptions of the 
variables $q,p$ going to $t,\omega$ ) as improvement over the Husimi $Q$ function and the 
Spectrogram $P_{SP}(t,\omega)$ respectively. 
       
In Sec. 2 we derive our basic result on the best possible positive normalized probability distribution 
closest to $W$. In Sec. 3 we solve the corresponding variational problem when additional rotationally 
invariant constraints in phase space are added.In the particular examples 
considered in this paper these additional constraints enable reproducing the 
correct uncertainty product for position and momentum. In Sec. 4 we calculate the two optimal distributions 
explicitly in the case of the generalized coherent states of quantum optics and compare them numerically 
with the Wigner and Husimi distributions in table 1 and Figs. 1 to 4.The results bring out not only that 
the optimal distributions are much closer to the Wigner distribution than the Husimi $Q$ function but 
also that the marginals of the optimal distributions are much closer to the true position probability 
density than those of the Husimi function.In Sec. 5 we outline a more ambitious problem of finding the 
positive normalized distribution closest to the Wigner function which reproduces both the position 
and momentum probabilities of quantum mechanics exactly.In Sec. 6 we summarise our conclusions. 

 {\bf 2. Positive joint probability distribution closest to the Wigner distribution and a measure 
of quantumness}.
Suppose we know $W$ through quantum tomography. We seek 
a criterion invariant under phase space rotations to define the positive definite phase space probability 
density `closest' to the $W$ function and with total phase space integral unity, as necessary for a probability 
interpretation. The criterion of `closeness' must be such that it gives back the $W$ function when that is 
positive definite. In $2N$ dimensional phase space, with units $\hbar = c = 1$, the Wigner function  
is given in terms of the density operator $ \rho $ ,

\begin{eqnarray}
 W(\vec q,\vec p)= \frac{1} {(2 \pi)^N} \int d\vec y \exp (i \vec p. \vec y) 
  \langle \vec q-\vec  y/2 |\rho  |\vec q+ \vec y/2 \rangle \nonumber \\
= \frac{1} {(2 \pi)^{2N}} \int d\vec \xi \int d\vec \eta \: Tr \rho \exp (i \vec \xi . (\vec q_{op} -\vec q ) + \nonumber \\
i \vec \eta . (\vec p_{op} -\vec p ) ) ,
\end{eqnarray}
where time dependence of the density operator and the Wigner function have been suppressed, $\vec q_{op},\vec p_{op} $ 
denote the position and momentum operators and the last equation facilitates discussion of rotation properties 
in phase space.In quantum optics, 
 
\begin{equation}
 \vec q _{op} = (\vec a +\vec a^{\dagger})/\sqrt {2},\: \vec p_{op}=-i (\vec a-\vec a ^{\dagger})/\sqrt {2}.
\end{equation}

We vary $P(\vec q , \vec p)$ so as to minimise,
\begin{equation}
 \sigma ^2 = \frac{\int d \vec q \int d \vec p \: (P(\vec q , \vec p)-W(\vec q , \vec p))^2 } 
{\int d\vec q \int d\vec p \: W(\vec q , \vec p)^2 } 
\end{equation}
(which is just the mean of the square of the fractional deviation $(P-W)/W $ with the weight function $W^2$ ), subject to the 
constraints, 
\begin{equation}
 \int d\vec q \int d\vec p \: P(\vec q , \vec p) =1 ; \:P(\vec q , \vec p) \geq 0 .
\end{equation}
We use Lagrange's method of undetermined multipliers modified to incorporate inequality constraints. The above 
normalization constraint is equivalent to
\begin{equation}
 \int d\vec q \int d\vec p \: (P(\vec q , \vec p)-W(\vec q , \vec p)) =0 ,
\end{equation}
and the expression for $\sigma^2$, using Moyal's well known result for phase space integral of $W^2$ \cite{Wigner} 
simplifies ,for pure states, to 
\begin{equation}
 \sigma ^2 
= (2 \pi)^N \int d\vec q \int d\vec p \: (P(\vec q , \vec p)-W(\vec q , \vec p))^2 .
\end{equation}
{\bf Remark }. For impure states, the factor $(2 \pi)^N $ on the right-hand side must be replaced 
by $(2 \pi)^N / Tr (\rho^2 )$.

This leads to the Lagrangian,
\begin{eqnarray}
 L=\int d \vec q \int d \vec p \:(P(\vec q , \vec p)-W(\vec q , \vec p))^2  \nonumber \\
+ 2 c \int d\vec q \int d\vec p \: (P(\vec q , \vec p)-W(\vec q , \vec p)) ,
\end{eqnarray}
where c is the Lagrange multiplier.
Following a method used widely by Martin to incorporate inequality constraints \cite{Martin}, 
we prove by direct subtraction that $ \sigma ^2$ has a global minimum when we choose 
$ P(\vec q , \vec p)= P_{min}(\vec q , \vec p)$, where,
\begin{equation}
   P_{min}(\vec q , \vec p)= P_0(\vec q , \vec p) \:\theta ( P_0(\vec q , \vec p) ),
\end{equation}
where $\theta (x)$ is the Heaviside $\theta $ function, being unity when the argument is positive and zero 
otherwise, and
 \begin{equation}
  P_0(\vec q , \vec p)= W(\vec q , \vec p)-c  .
 \end{equation}
 Denoting by $L$ and $L_{min}$ respectively the values of the Lagrangian for an arbitrary $ P(\vec q , \vec p) $
satisfying the constraints, and by $ P_{min}(\vec q , \vec p) $ , we obtain,

\begin{eqnarray}
 L - L_{min} = \int _{P_0 \geq 0} (P-P_0)^2 d \vec q d \vec p  \nonumber \\
+ \int _{P_0 \leq 0} (P^2 -2P P_0) d \vec q d \vec p \geq 0 , 
\end{eqnarray}
since each of the two integrands is non-negative.
We complete the proof by showing the existence and uniqueness of a constant $c$ satisfying the normalization constraint,
\begin{equation}
 \int _{W(\vec q , \vec p)-c\geq 0} (W(\vec q , \vec p)-c) d \vec q d \vec p = 1 .
\end{equation}
First, if $W$ is non-negative, $c=0$ is the unique solution, and gives $ \sigma ^2 =0$. Suppose now that $W$ is 
negative in some regions of phase space. The left-hand side integral is then $\geq 1 $ for $c \leq 0$, decreases 
monotonically as c increases to positive values until it equals $0$ when $c= max_{\vec q,\vec p} W(\vec q,\vec p)$.
Hence there is a unique solution for $c$ in the interval $[0, max_{\vec q,\vec p} W(\vec q,\vec p)]$.  
Using this value of $c$ we compute the optimum phase space probability distribution as well as the minimum value 
of $ \sigma ^2$, an index of quantumness of the state . 

 {\bf 3. Incorporating additional rotationally invariant constraints in phase space }. The variational 
method outlined above is invariant under phase space rotations. Can we incorporate other quantum constraints 
preserving such invariance?
In addition to the phase space volume, the surface of the sphere with centre $\vec q_{cl} ,\vec p_{cl} $,
$$(\vec q -\vec q_{cl})^2 + (\vec p -\vec p_{cl})^2  =x $$ 
is an invariant under rotations in phase space ,and hence may be used as an additional constraint.
With a view towards imposing the correct sum of quantum dispersions
 $(\Delta \vec q)^2 + (\Delta \vec p)^2$ on the variational phase space density, 
we choose $\vec q_{cl} ,\vec p_{cl} $ as the quantum 
expectation values of $\vec q_{op} ,\vec p_{op} $.
Further, if $W$ remains positive in the region $x \geq x_{max}$, 
we may choose $P(\vec q , \vec p)= W(\vec q , \vec p)$ in that region, and  
for sufficiently large $x_{max}$ ,still find a solution $P(\vec q , \vec p)$ that 
minimises $ \sigma ^2$ under the positivity constraint $P(\vec q , \vec p) \geq 0$,
the normalisation constraint,
\begin{equation}
\int \int _{x \leq x_{max}} d\vec q d\vec p \: (P(\vec q , \vec p)-W(\vec q , \vec p)) =0 ,
\end{equation}
and the additional constraint,
\begin{equation}
\int \int _{x \leq x_{max}} d\vec q  d\vec p \: (P(\vec q , \vec p)-W(\vec q , \vec p)) x =0 .
\end{equation}
The last equation imposes the sum of quantum dispersions $(\Delta \vec q)^2 + (\Delta \vec p)^2$ on $P$ 
since the Wigner function obeys that constraint.
We then prove as before that the solution minimising $ \sigma ^2$ is, for $x \leq x_{max} $
\begin{equation}
   P_{min1}(\vec q , \vec p)= P_{01}(\vec q , \vec p) \:\theta ( P_{01}(\vec q , \vec p) ),
\end{equation}
where 
\begin{equation}
  P_{01}(\vec q , \vec p)= W(\vec q , \vec p)-c -x d ,
\end{equation}
provided that constants $c,d$ are found satisfying the two equality constraints given above.

 {\bf 4. Optimum positive joint probability distributions and Husimi distribution for generalized 
coherent states }. The Husimi $Q$ function in $2N$-dimensional phase space is,
\begin{equation}
 Q(\vec q,\vec p)= (2 \pi )^{-N} \langle \vec \alpha \vert \rho \vert \vec \alpha \rangle 
\end{equation}
where $\vert \alpha \rangle $ are the coherent states,
\begin{equation}
\vec a \vert \vec \alpha \rangle = \vec \alpha \vert \vec \alpha \rangle \:,\vec \alpha = (\vec q +i \vec p )/ \sqrt{2}., ,
\end{equation}
Generalized coherent states \cite{Roy-Singh} are displaced excited eigen state solutions of the time dependent 
Schr$\ddot{o}$dinger equation for the one dimensional oscillator whose probability density packets 
move classically with shape unchanged, and have uncertainty product $\Delta q \Delta p =n+1/2$, 
\begin{eqnarray}
 \langle q \vert \psi (t) \rangle = \langle q-q_{cl}(\tau) \vert n \rangle \exp ( {-i (n+1/2)\tau } ) \nonumber \\
\exp ( {i \dot{q}_{cl}(\tau) (q-1/2 \dot{q}_{cl}(\tau) ) )} ,
\end{eqnarray}
where, $\vert n \rangle$ is the $n$-th excited state and $q_{cl}$ has classical motion 
\begin{equation}
\tau =\omega t,\: q_{cl}(\tau)= A \cos (\tau + \phi). 
\end{equation}
The quantum expectation values for position and momentum operators are,
\begin{equation}
 \langle q_{op} \rangle = q_{cl} (\tau),\: \langle p_{op} \rangle = \dot {q}_{cl} (\tau)\equiv p_{cl}.
\end{equation}
Wigner functions and Husimi functions can be seen to depend on $q,p$ only through the combination,
\begin{equation}
x= ( q - q_{cl})^2 + ( p - p_{cl})^2 .
\end{equation}
For $n=0$ the optimum phase space probability density is just the Wigner function which is positive definite.
For $n=1,2$ the $W_n (q,p)$ and $Q_n (q,p)$ functions are given by,
\begin{eqnarray}
W_1 =(2/\pi)(x-1/2)\exp{(-x)}, \nonumber \\
Q_1 = (x/(4 \pi))\exp{(-x/2)},
\end{eqnarray}
\begin{eqnarray}
W_2 = (2/\pi)((x-1)^2-1/2)\exp{(-x)} ,\nonumber \\
Q_2 = (x^2/(16 \pi))\exp{(-x/2)}.
\end{eqnarray}

 We have numerically evaluated the optimum phase space probability distribution
$P_{min} $ of Sec.2 with only positivity and normalization constraint, and $P_{min1} $ of Sec.3 
with the additional constraint of the correct $\Delta q \Delta p $ for the generalized coherent states 
with $n=1$ and $n=2$. We have also evaluated the corresponding Husimi $Q$ distributions.
We compared the optimum $P_{min},P_{min1} $ with $W, Q$ distributions in Figs. 1,2. We also compared the 
corresponding position probability densities in Figs. 3,4. Both of the optima $P_{min},P_{min1} $ show 
a big improvement over the Husimi function, as is obvious qualitatively from the figures, and 
quantitatively from the $\sigma ^2 $ values listed in the table.
 
\begin{figure}[ht]
\begin{center}
 \includegraphics[width= 1.\columnwidth]{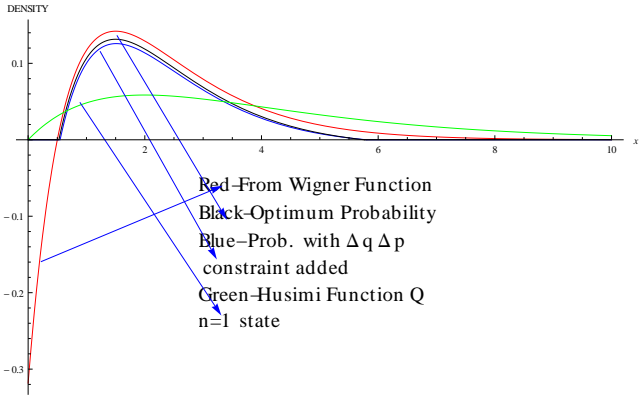}
\caption{ For the n=1 coherent state , the optimum phase space probability distributions 
with only normalization constraint (black), and including additional constraints fixing $\Delta q \Delta  p$ 
(blue) are compared with the Wigner (red) and Husimi (green) distributions as a function of
$x= ( q - q_{cl})^2 + ( p - p_{cl})^2$. The optimum and Husimi distributions have 
 $ \sigma ^2 =0.277049 $, and $0.509259 $ respectively.} 
\label{fig:phase_space_probability_n=1}
\end{center}
\end{figure}

\begin{figure}[ht]
\begin{center}
 \includegraphics[width=1.\columnwidth]{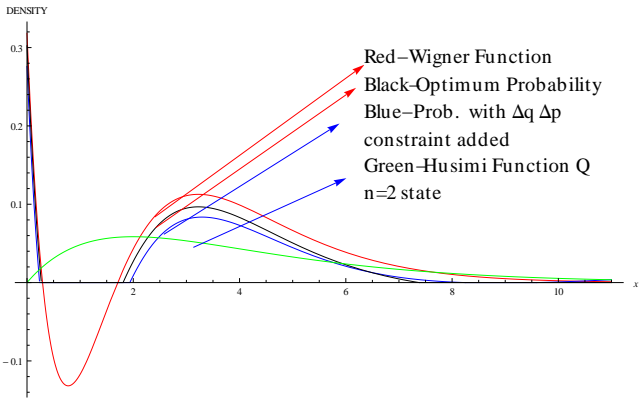}
\caption{The same plots as in Fig.1 for the n=2 coherent state .
The optimum and Husimi distributions have  $ \sigma ^2 =0.268084$, and $0.64429 $ respectively.}
\label{fig:phase_space_probability_n=2}
\end{center}
\end{figure}

\begin{figure}[ht]
\begin{center}
 \includegraphics[width=1.\columnwidth]{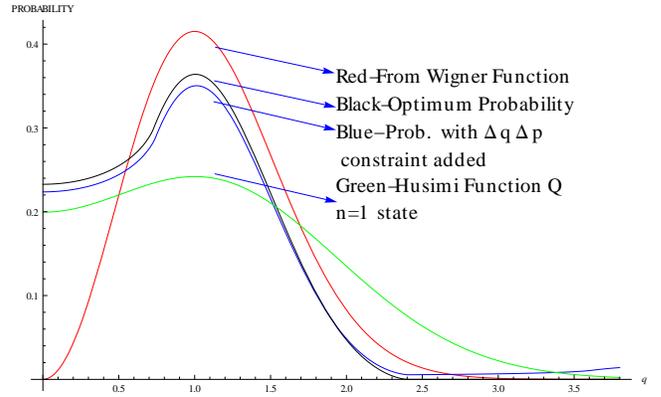}
\caption{For the n=1 coherent state , the position probabilities calculated from the 
optimum joint probabilities with only normalization constraint (black), and with additional constraints 
fixing $\Delta q \Delta  p$ (blue) are seen to be closer to the true probability (given by the Wigner distribution
(red)) than the Husimi distribution result (green).}
\label{fig:position_probability_n=1}
\end{center}
\end{figure}

\begin{figure}[ht]
\begin{center}
 \includegraphics[width=1.\columnwidth]{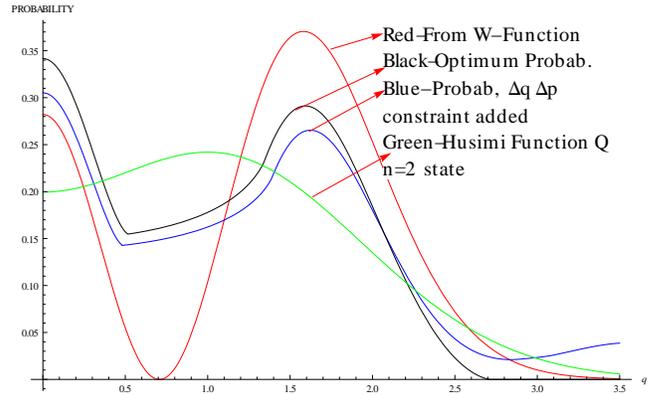}
\caption{Same plots as in Fig.3, for the n=2 coherent state of the oscillator.}
\label{fig:position_probability_n=2}
\end{center}
\end{figure}

\begin{center}
\begin{table}[hbt]
 \caption {Husimi Function versus Optimum Probability Distributions; $\sigma^2$ is the mean square fractional 
deviation from the Wigner distribution.}
 \begin{tabular}{|c|c|c|c|c|c|c|} \hline \hline
  {\em State}&{\em Husimi}&\multicolumn{5}{c|}{\em Optimum Probability }\\
 &{\em Function}&\multicolumn{5}{c|}{\em Density}\\ \cline{2-7}
  &$\sigma ^2$ & $\sigma ^2$ &$c$ &$d$ & $x_{max }$& $\Delta q \Delta p$ \\ \hline
n=1&.5093&.2770&.01053&0&$\infty $&1.108 \\ \cline{3-7}
$\Delta q \Delta p =3/2$& &.2877&.01837&-.0014&18&3/2 \\ \hline
n=2&.6443&.2681&.01595&0&$\infty $ &1.722 \\ \cline{3-7}
$\Delta q \Delta p =5/2$& &.3223&.04235&-.00408&15&5/2 \\ \hline \hline
 \end{tabular}
\end{table}
\end{center}

 {\bf 5. Optimum Positive Phase Space Densities Reproducing $N+1$ Quantum Marginals }. Cohen and 
Zaparovanny \cite{Cohen-Z} constructed the most general 
positive phase space densities reproducing two marginals of $W$, viz. quantum probability densities 
of $\vec q$ and $\vec p$ . In $2N$-dimensional phase space, with $N\geq 2$, Roy and Singh \cite{Roy-Singh99} noted 
that in fact $N+1$ marginals of $W$ (e.g. for $N=2$, probability densities 
of $(q_1, q_2), (p_1, q_2), (p_1, p_2)$) can be reproduced with positive densities; they conjectured 
that no more than $N+1$ marginals can be so reproduced for arbitrary quantum states, the $``N+1''$ marginal theorem.
This was proved later using an extension of Bell inequalities \cite{Bell} to phase space by 
Auberson et al\cite{Auberson}, who also derived the most general positive phase space density reproducing 
$N+1$ marginals; that density is non-unique since it contains an abitrarily specifiable phase space function.  
Among the continuous infinity of positive phase space densities reproducing $N+1$ marginals which one is closest 
to the Wigner Function ? Our method gives a straight forward answer; we give the variational answer 
explicitly for $N=2$, and indicate briefly the generalization to $N \geq 2 $.   
Find the phase space density $ P( q, p)$ obeying positivity, minimum mean square fractional 
deviation from the Wigner distribution , reproducing the quantum probability densities of 
$q$, and  $p$ . Vary $ P ( q, p ) $ to minimise the Lagrangian, 
\begin {equation}
 L = \int  [(P-W)^2 + (2 \lambda (q) + 2\mu (p) ) (P-W) ] d q d p ,
\end {equation}
subject to the constraints,
\begin {equation}
 \int (P-W)  d p = 0, \:\int (P-W) d q  = 0 ,\:P (\vec q,\vec p ) \geq 0 .
\end {equation}
$L$ is minimised if we choose for $P$, 
the function $P_0$ that makes $L$ stationary whenever $P_0$ is positive, and zero otherwise:
\begin {equation}
 P_{min}= P_0 \theta (P_0), \:P_0 \equiv W -\lambda (q) -\mu (p), 
\end {equation}
where the multipliers $\lambda (q),\mu (p) $ are determined from the constraints. As in Sec. 2, we prove 
by direct subtraction that $ L-L_{min} \geq 0 $, the only change being the new choice of 
$P_0 \equiv W -\lambda (q) -\mu (p)$.
The constraints yield a pair of coupled integral equations to determine $\lambda (q) , \mu (p) $:
\begin{eqnarray}
 \int _ {P_0 \geq 0} (\lambda (q) + \mu (p) ) dp = - \int _{P_0 \leq 0} W(q,p) dp ,\nonumber \\
\int _ {P_0 \geq 0} (\lambda (q) + \mu (p) ) dq = - \int _{P_0 \leq 0} W(q,p) dq \>,
\end{eqnarray}
which complete evaluation of the optimum phase space density.For $N\geq 2$, the positivity constraint is 
supplemented by $N+1$ marginal constraints, which can, for example, be chosen to be the series of 
probability densities of $(q_1, q_2,..q_n), (p_1, q_2,..q_n),.. (p_1, p_2,..p_n)$, in which each member is 
obtained by replacing in the previous set one co-rdinate by its conjugate momentum.The optimal phase space density is 
again constructed by a Lagrange multiplier method which will now involve $N+1$ Lagrange multiplier functions.  

{\bf 6. Conclusion.} We have proposed a general method to find the positive phase space distribution closest 
to the Wigner distribution that can be used in quantum optics as well as in time frequency analysis. A 
measure of quantumness  emerges. Qualitative and quantitative improvement  with 
respect to the Husimi function is seen explicitly; e.g. for the generalized coherent states, 
the optimum and Husimi distributions have respectively, for $  n=1\:$, $\sigma ^2=.277049$, and $0.509259 $, 
for $ n=2\: $, $ \sigma ^2 =0.268084$, and $0.64429 $ . Similar improvements 
are expected in time frequency analysis.In $2N$-dimensional phase space the optimum positive density reproducing 
$N+1$ marginals can be evaluated.

 The authors thank G. Auberson and 
G. Mahoux for the remark on impure states following Eq.(12), and the referee for many useful suggestions. SMR thanks 
Sumit Das for a remark many years ago on invariance under canonical transformations, Aditi Sen De, Ashok Sen and 
R. Gopakumar for discussions, and the Indian National Science Academy for financial support. 
A preliminary outline of this work was presented at a recent lecture  \cite{SMR}.


\end{document}